\documentclass{article}

\usepackage[preprint]{neurips_2024}

\usepackage[utf8]{inputenc} 
\usepackage[T1]{fontenc}    
\usepackage{hyperref}       
\usepackage{url}            
\usepackage{booktabs}       
\usepackage{amsfonts}       
\usepackage{nicefrac}       
\usepackage{microtype}      
\usepackage{xcolor}         
\usepackage{threeparttable}
\usepackage{amsmath}
\usepackage{adjustbox}
\usepackage{lscape}
\usepackage{tabularx}
\usepackage{booktabs}
\usepackage{rotating}

\title{Impact of Major Health Events on Pharmaceutical Stocks: A Comprehensive Analysis Using Macroeconomic and Market Indicators}

\author{%
  Morteza Maleki \\
  Georgia Institute of Technology,\\
  School of Industrial and Systems Engineering,\\
  Atlanta, Georgia, USA \\
  \texttt{mmaleki3@gatech.edu} \\
  \And
  SeyedAli Ghahari \\
  Purdue University, \\
  Department of Civil and Environmental Engineering,\\
  West Lafayette, USA \\
  \texttt{sghahari@purdue.edu} \\
}

\begin{document}

\maketitle

\begin{abstract}
This study investigates the impact of significant health events on pharmaceutical stock performance, employing a comprehensive analysis incorporating macroeconomic and market indicators. Using Ordinary Least Squares (OLS) regression, we evaluate the effects of thirteen major health events since 2000, including the Anthrax attacks, SARS outbreak, H1N1 pandemic, and COVID-19 pandemic, on the pharmaceutical sector. The analysis covers different phases of each event—beginning, peak, and ending—to capture their temporal influence on stock prices. Our findings reveal distinct patterns in stock performance, driven by market reactions to the initial news, peak impact, and eventual resolution of these crises. We also examine scenarios with and without key macroeconomic (MA) and market (MI) indicators to isolate their contributions. This detailed examination provides valuable insights for investors, policymakers, and stakeholders in understanding the interplay between major health events and health market dynamics, guiding better decision-making during future health-related disruptions.
\end{abstract}

\section{Introduction}

The pharmaceutical industry is uniquely sensitive to public major health events, which can significantly influence stock market performance. Historically, events such as pandemics, disease outbreaks, and significant medical advancements have led to substantial volatility in pharmaceutical stock prices. For instance, the SARS outbreak in 2003 and the H1N1 pandemic in 2009 both caused notable market disruptions, reflecting the industry's vulnerability to sudden health threats \cite{chen2009positive, wang2022pandemic}.

Macroeconomic indicators, such as GDP, inflation rates, and unemployment rates, also play a critical role in determining stock market dynamics. These factors can compound the effects of major health events on the pharmaceutical sector. Previous studies have demonstrated that economic downturns, coupled with health emergencies, can exacerbate stock market fluctuations \citep{wu2021does, tsai2015us}. This underscores the importance of a comprehensive analysis that includes both macroeconomic and market indicators to understand the full impact of major health events on pharmaceutical stocks.

Moreover, Market indicators, such as the S\&P 500 and NASDAQ indices, are crucial for contextualizing the performance of pharmaceutical stocks within the broader market environment. These indices provide a benchmark against which the specific impacts of major health events can be measured \cite{yahoo_nasdaq, yahoo_sp500}. Additionally, historical event data, such as the approval of new vaccines or the emergence of novel diseases, offer invaluable insights into how specific events influence investor behavior and stock prices.

This study aims to bridge the gap in existing literature by providing a detailed examination of how major health events affect pharmaceutical stocks, incorporating both macroeconomic and market indicators. By analyzing thirteen major health events over the past two decades (since 2000), we seek to uncover patterns and trends that can guide investors, policymakers, and industry stakeholders in making informed decisions during health emergencies.

\subsection{Background}

The relationship between public major health events and financial markets has been a subject of interest for researchers and practitioners alike.\cite{maleki2022social, healthcare12151458} During the SARS outbreak in 2003, for example, there was a significant decline in stock prices across various sectors, with the pharmaceutical industry experiencing mixed outcomes \citep{siu2004economic}. Similarly, the H1N1 pandemic in 2009 led to heightened market volatility, reflecting investor uncertainty and risk aversion \citep{velasquez2022emerging}.

Macroeconomic factors further complicate this relationship. Economic indicators such as GDP growth, inflation, and unemployment can influence investor confidence and market stability. For instance, periods of economic recession often see heightened sensitivity to external shocks, including major health events \citep{barro2008macroeconomic}. Conversely, economic booms may buffer the negative impacts of such events on stock prices.

\subsection{Purpose of the Study}

The primary objective of this study is to evaluate the impact of major major health events on pharmaceutical stock performance. By examining events such as the Anthrax attacks, SARS outbreak, and COVID-19 pandemic, we aim to understand how these crises influence stock prices over different phases—beginning, peak, and ending of the event. This temporal analysis allows us to capture the dynamic nature of market reactions to health emergencies.

We also aim to disentangle the effects of macroeconomic and market indicators from the direct impacts of major health events. By creating scenarios that include and exclude these indicators, we can isolate their contributions and provide a clearer picture of the factors driving stock market responses during health emergencies.

\subsection{Novelty and Significance}

This study makes several significant contributions to the existing literature. Firstly, by incorporating a wide range of macroeconomic and market indicators, as summarized in Table \ref{tab:all_variables}, we provide a comprehensive analysis of the factors influencing pharmaceutical stock performance during major health events. This multi-faceted approach allows us to account for the complex interplay between health events and economic conditions.

Secondly, our analysis of thirteen major health events over the past two decades, as summarized in \ref{tab:events} offers a robust dataset for examining patterns and trends. This extensive temporal coverage enhances the reliability of our findings and provides valuable historical context for understanding current and future market dynamics.

Lastly, the inclusion of different phases of each event—beginning, peak, and ending —provides a nuanced understanding of market reactions. This temporal breakdown allows us to capture the immediate, intermediate, and long-term impacts of major health events on pharmaceutical stocks, offering deeper insights for investors and policymakers.

\subsection{Analysis Framework}

In this study, the impact of 13 important major health events events between 2000 and 2023 on select pharmaceutical company stocks, as summarized in Table \ref{tab:tickers_companies} has been analyzed, while controlling for an extensive list of Macroeconomic variables and Market related factors.

\begin{table}
\caption{Ticker Symbols and Corresponding Companies}
\label{tab:tickers_companies}
\centering
\begin{tabularx}{\linewidth}{p{2cm}p{4cm}p{2cm}p{4cm}}
\toprule
\textbf{Ticker} & \textbf{Company} & \textbf{Ticker} & \textbf{Company} \\
\midrule
JNJ & Johnson \& Johnson & LLY & Eli Lilly \\
PFE & Pfizer & GSK & GlaxoSmithKline \\
MRK & Merck & NVO & Novo Nordisk \\
ABBV & AbbVie & AMGN & Amgen \\
TMO & Thermo Fisher Scientific & AZN & AstraZeneca \\
GILD & Gilead Sciences & & \\
\bottomrule
\end{tabularx}
\end{table}

\begin{figure*}
\centering
\includegraphics[width=14cm]{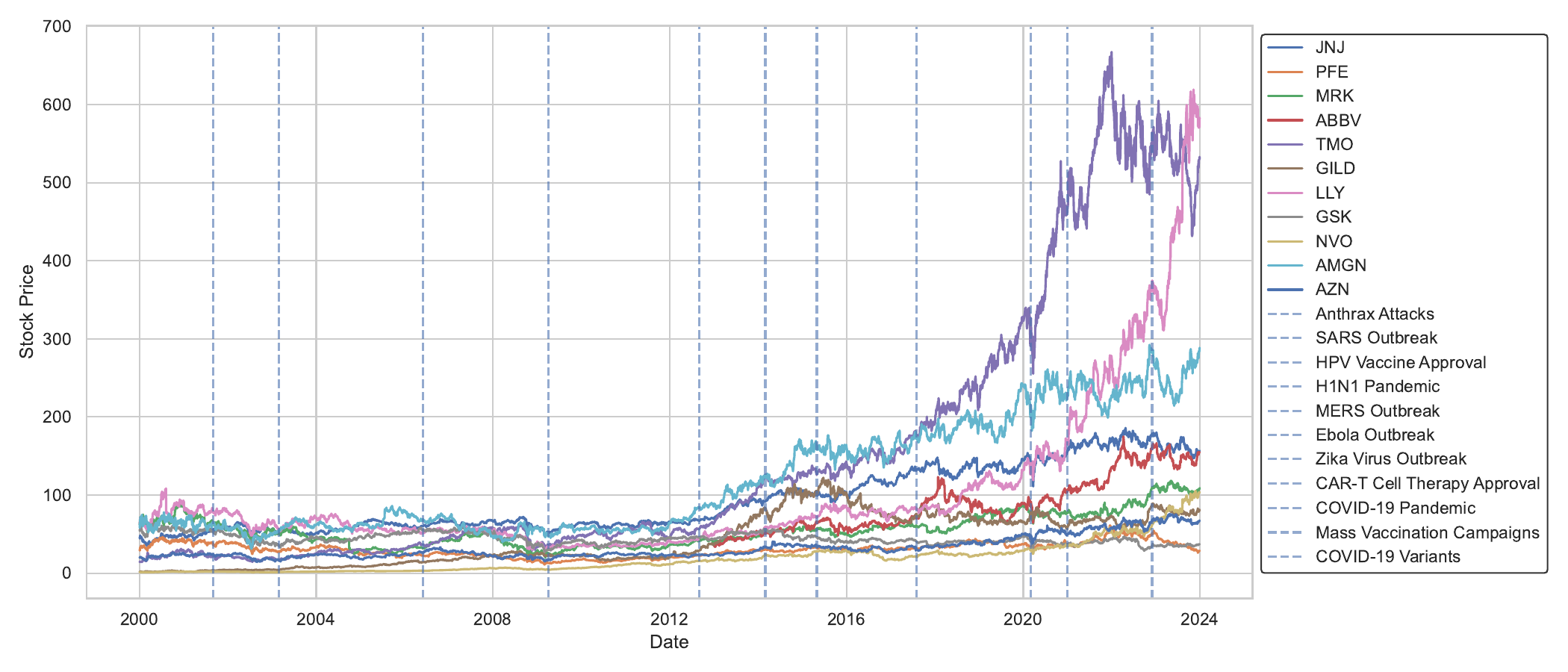}
\caption{This figure shows the time series of closing prices for major pharmaceutical companies (JNJ, PFE, MRK, ABBV, TMO, GILD, LLY, GSK, NVO, AMGN, AZN) from 2000 to 2023. Significant health events are marked with dashed vertical lines, including Anthrax Attacks (2001), SARS Outbreak (2003), HPV Vaccine Approval (2006), H1N1 Pandemic (2009), MERS Outbreak (2012), Ebola Outbreak (2014), Zika Virus Outbreak (2016), CAR-T Cell Therapy Approval (2017), COVID-19 Pandemic (2020), Mass Vaccination Campaigns (2021), and COVID-19 Variants (2022). Each event's period (start, peak, and end) is indicated to highlight their impact on stock prices.\label{fig:all_timeseries}}
\end{figure*}

Figure \ref{fig:all_timeseries}, illustrates the closing stock price value for the select pharma company stocks over this period of time in addition to the major health events occurrences in the same period.

\section{Literature Review}

Over the past two decades, extensive research has been conducted to understand the factors influencing the stock prices of pharmaceutical companies. This section reviews the relevant literature, focusing on the impacts of macroeconomic variables, market trends, and major health events.

\subsection{Macroeconomic Variables and Stock Prices}

The relationship between macroeconomic variables and stock prices has been a subject of considerable research \citep{ibrahim1999macroeconomic, hunjra2014impact}. Ibrahim et al. analyzed the impact of various macroeconomic indicators, including GDP, inflation, and unemployment, on the stock prices of pharmaceutical companies. The study found significant correlations between these variables and stock performance, suggesting that macroeconomic conditions play a crucial role in shaping investor perceptions and market dynamics.

Similarly, Otoo et al. \citep{otoo1999consumer} examined the influence of consumer sentiment and interest rates on pharmaceutical stocks. Their findings indicate that positive consumer sentiment and lower interest rates generally lead to higher stock prices, as they signal economic stability and growth prospects.

\subsection{Market Trends and Sectoral Analysis}

Market trends and broader financial market dynamics also significantly impact pharmaceutical stock prices. Karamehic et al. \citep{karamehic2013financial} conducted a comprehensive review of financial markets and the pharmaceutical industry. They highlighted the sector's sensitivity to market indices such as the S\&P 500 and NASDAQ, noting that fluctuations in these indices often lead to corresponding changes in pharmaceutical stock prices.

Research by Mironiuc et al. \citep{mironiuc2022financial} further emphasized the sector's volatility, particularly during periods of market uncertainty. Their study on the COVID-19 pandemic revealed that pharmaceutical stocks experienced significant volatility, reflecting the market's response to the crisis and the heightened demand for healthcare solutions.

\subsection{major health events and Pharmaceutical Stocks}

major health events, including pandemics and epidemics, have profound impacts on the pharmaceutical sector. Velasquez at al. \cite{velasquez2022emerging} explored the economic impact of various major health events on pharmaceutical stocks. They found that events like the SARS outbreak in 2003 and the H1N1 pandemic in 2009 led to increased stock volatility as companies raced to develop treatments and vaccines.

Additionally, the COVID-19 pandemic has been the most studied health crisis in recent history due to its global scale and impact. Research by Al-awadhi, Ashraf, and Zhang et al. \cite{al-awadhi2020, ashraf2020, zhang2020} showed that the pandemic caused unprecedented volatility in pharmaceutical stocks, with companies involved in vaccine development and production seeing substantial gains. This aligns with earlier findings by Denielsson et al. \cite{danielsson2018learning}, who noted that crises events often lead to short-term gains for certain pharmaceutical companies, followed by periods of high volatility.

\subsection{Predictive Modeling in Financial Markets}

Predictive modeling has become an essential tool in financial markets for forecasting stock prices and understanding the impacts of various factors. Recent advancements in machine learning and econometric modeling have enabled more accurate predictions. For instance, studies using Ridge and Lasso regressions have demonstrated their effectiveness in handling multi-collinearity and selecting significant predictors \citep{hoerl1970ridge, tibshirani1996lasso}.

Moreover, in the context of the pharmaceutical sector, predictive models have been used to forecast stock price movements based on macroeconomic indicators and market trends. Research by Tibshirani \citep{tibshirani1996lasso} introduced the Lasso regression method, which has since been widely adopted for its ability to produce sparse models that are easier to interpret. This technique is particularly useful in financial modeling, where it is important to identify the most influential variables among a large set of predictors.

\subsection{Stock Market Prediction using Macroeconomic Factors}

The literature on stock market prediction and the impact of macroeconomic factors on stock prices is extensive. Studies have employed various machine learning algorithms to forecast stock index movements with significant accuracy. Milosevic demonstrated the effectiveness of hybrid machine learning algorithms in predicting the daily return direction of the stock market \cite{milosevic2016equity}. Similarly, Kumar and Thenmozhi (2006) compared support vector machines and random forest algorithms for forecasting stock index movements, highlighting the potential of these techniques in financial prediction \cite{kumar2006forecasting}. Furthermore, Akaev and Sadovnichii (2020) used a mathematical model of Hyman Minsky's Theory of Financial Instability to forecast a cyclical downturn in the US economy, underscoring the relevance of theoretical models in economic forecasting \cite{akaev2021information}. Borio et al. (2019) also contributed to this field by comparing the predictive power of the financial cycle and the term spread in forecasting recessions \cite{borio2019predicting}. These studies collectively emphasize the importance of leveraging advanced predictive techniques and theoretical models to enhance the accuracy of financial market forecasts.

\section{Materials and Methods}

This section outlines the materials and methods used in the study, including data sources, data processing, aggregation techniques, and the methodologies employed for the analysis. The aim is to provide a comprehensive understanding of the steps involved in the research process.

\begin{table*}[htbp]
\centering
\caption{List of variables and their descriptions used in the study. The table includes stock data (e.g., average open price, high price, low price), market data (e.g., S\&P 500 Index, NASDAQ Index), and macroeconomic data (e.g., GDP, inflation rate, unemployment rate). Each variable is accompanied by its code, category, data source, timeframe, and frequency.}
\label{tab:all_variables}
\begin{tabularx}{\linewidth}{|p{1.5cm}|p{3.5cm}|p{3cm}|p{2cm}|p{1.8cm}|}
    \hline
    \textbf{Code} & \textbf{Category} & \textbf{Data Source} & \textbf{Timeframe} & \textbf{Frequency} \\
    \hline
    ST-OM & Stock Data & Yahoo Finance & 2000-2023 & Daily \\
    ST-HM & Stock Data & Yahoo Finance & 2000-2023 & Daily \\
    ST-LM & Stock Data & Yahoo Finance & 2000-2023 & Daily \\
    ST-CM & Stock Data & Yahoo Finance & 2000-2023 & Daily \\
    ST-AM & Stock Data & Yahoo Finance & 2000-2023 & Daily \\
    ST-VM & Stock Data & Yahoo Finance & 2000-2023 & Daily \\
    ST-VS & Stock Data & Yahoo Finance & 2000-2023 & Daily \\
    \hline
    MI-SP & Market Data & Yahoo Finance & 2000-2023 & Daily \\
    MI-NS & Market Data & Yahoo Finance & 2000-2023 & Daily \\
    \hline
    MA-GP & Macroeconomic Data & FRED & 2000-2023 & Quarterly \\
    MA-IF & Macroeconomic Data & FRED & 2000-2023 & Monthly \\
    MA-UR & Macroeconomic Data & FRED & 2000-2023 & Monthly \\
    MA-FF & Macroeconomic Data & FRED & 2000-2023 & Monthly \\
    MA-CS & Macroeconomic Data & FRED & 2000-2023 & Monthly \\
    MA-IP & Macroeconomic Data & FRED & 2000-2023 & Monthly \\
    MA-M2 & Macroeconomic Data & FRED & 2000-2023 & Monthly \\
    MA-MR & Macroeconomic Data & FRED & 2000-2023 & Weekly \\
    MA-TR & Macroeconomic Data & FRED & 2000-2023 & Monthly \\
    MA-NP & Macroeconomic Data & FRED & 2000-2023 & Monthly \\
    MA-HS & Macroeconomic Data & FRED & 2000-2023 & Monthly \\
    MA-RS & Macroeconomic Data & FRED & 2000-2023 & Monthly \\
    MA-PC & Macroeconomic Data & FRED & 2000-2023 & Monthly \\
    MA-PP & Macroeconomic Data & FRED & 2000-2023 & Monthly \\
    MA-TB & Macroeconomic Data & FRED & 2000-2023 & Monthly \\
    MA-CP & Macroeconomic Data & FRED & 2000-2023 & Quarterly \\
    MA-GD & Macroeconomic Data & FRED & 2000-2023 & Quarterly \\
    MA-ER & Macroeconomic Data & FRED & 2000-2023 & Monthly \\
    MA-GI & Macroeconomic Data & FRED & 2000-2023 & Quarterly \\
    MA-PS & Macroeconomic Data & FRED & 2000-2023 & Monthly \\
    MA-CC & Macroeconomic Data & FRED & 2000-2023 & Monthly \\
    MA-RI & Macroeconomic Data & FRED & 2000-2023 & Monthly \\
    MA-CU & Macroeconomic Data & FRED & 2000-2023 & Monthly \\
    MA-IE & Macroeconomic Data & FRED & 2000-2023 & Monthly \\
    \hline
    \end{tabularx}
\end{table*}

\subsection{Data and Data Sources}

The study utilizes a range of data sources to capture the various factors affecting pharmaceutical stock prices. The data is categorized into four main groups: stock data \cite{yahoo_finance}, market data \cite{yahoo_nasdaq, yahoo_sp500}, macroeconomic data \cite{FRED_API}, and event data \cite{cdc, fda, who}.

\subsubsection{Stock Data}

Stock data for pharmaceutical companies is sourced from Yahoo Finance \cite{yahoo_finance} and is summarized in Table \ref{tab:all_variables}. For the purposes of this study, since all Stock Data variables are highly inter-correlated, and to avoid multi-collinearity in regression analyses, only ST-CM (Average Closing Price across all select pharma companies) was used in the regression analyses and as a target variable.

\subsubsection{Market Data}

Market data includes indices that are indicative of the overall market performance. The markets leveraged in this analysis are providing below, and additional information is provided in Table \ref{tab:all_variables}. The data was sourced from Yahoo Finance S\&P 500 and NASDAQ Indices \cite{yahoo_nasdaq, yahoo_sp500}.

\begin{itemize}
    \item \textbf{S\&P 500 Index (MI-SP)}
    \item \textbf{NASDAQ Index (MI-NS)}
\end{itemize}

\subsubsection{Macroeconomic Data}

Macroeconomic data is collected from the Federal Reserve Economic Data (FRED) API \cite{FRED_API} and includes variables that influence the broader economy. The full list of the macroeconomic variables used in the study is provided in \ref{tab:all_variables}.

\subsubsection{Event Data}

Event data includes significant health-related events that impacted the pharmaceutical sector from 2000 - 2023 sourced from \cite{cdc, who, fda}. In this study, we added dummy variables corresponding to the health events in our dataset, and considered the health event's initial emergence impact for about 60 days divided into beginning, peak, and ending of the emergence of the event. This should not be confused with the peak and ending of the major health event in general which might have occurred months or years later. For instance, COVID-19 pandemic has been assumed to have begun on 2023-03-01, therefore, the beginning phase of COVID-19 pandemic has been considered all the dates between 2020-03-01 and 2020-03-31 (30 days), the peak for the COVID-19 shock to the market emergence is considered to be between 2020-03-15 and 2020-04-15 (30 days), and the period between 2020-04-01 until 2020-04-30. Same logic applies to the peak and ending phases of the \textbf{emergence} of other major health events. Table \ref{tab:events} summarizes the major health events considered in this study and their time-frame for the beginning, peak, and ending of each event's emergence and shock to the market.

\begin{table*}[htbp]
\centering
\caption{This table provides a detailed overview of significant health-related events considered in the study, including anthrax attacks, SARS outbreak, HPV vaccine approval, H1N1 pandemic, MERS outbreak, Ebola outbreak, Zika virus outbreak, CAR-T cell therapy approval, COVID-19 pandemic, mass COVID-19 vaccinations, emergence of COVID-19 variants, long COVID research, and advancements in AI drug development.}
\label{tab:events}
\begin{tabularx}{\linewidth}{|p{1.2cm}|p{4.5cm}|p{2cm}|p{2cm}|p{2cm}|}
\toprule
\textbf{Code} & \textbf{Event Name} & \textbf{Beginning} & \textbf{Peak} & \textbf{Ending} \\
\midrule
EV-AA & Anthrax Attacks & 2001-10-01 & 2001-10-15 & 2001-10-30 \\
EV-SO & SARS Outbreak & 2003-03-01 & 2003-03-15 & 2003-03-30 \\
EV-HV & HPV Vaccine Approval & 2006-06-01 & 2006-06-15 & 2006-06-30 \\
EV-HP & H1N1 Pandemic & 2009-04-01 & 2009-04-15 & 2009-04-30 \\
EV-MO & MERS Outbreak & 2012-05-01 & 2012-05-15 & 2012-05-30 \\
EV-EO & Ebola Outbreak & 2014-08-01 & 2014-08-15 & 2014-08-30 \\
EV-ZV & Zika Virus Outbreak & 2016-02-01 & 2016-02-15 & 2016-02-28 \\
EV-CT & CAR-T Cell Therapy Approval & 2017-08-01 & 2017-08-15 & 2017-08-30 \\
EV-CP & COVID-19 Pandemic & 2020-03-01 & 2020-03-15 & 2020-03-30 \\
EV-MV & Mass Vaccination Campaigns & 2020-12-01 & 2020-12-15 & 2020-12-30 \\
EV-CV & COVID-19 Variants & 2021-12-01 & 2021-12-15 & 2021-12-30 \\
EV-LC & Long COVID Research & 2023-03-01 & 2023-03-15 & 2023-03-30 \\
\bottomrule
\end{tabularx}
\end{table*}

\subsection{Data Processing and Aggregation}

The data processing involved several steps to ensure the dataset was ready for analysis:

\subsubsection{Data Cleaning}

The initial data contained missing values and inconsistencies that were addressed through:
\begin{itemize}
    \item Removing duplicates
    \item Imputing missing values using median values where applicable
    \item Normalizing date formats to ensure consistency
\end{itemize}

\subsubsection{Aggregation Techniques}

For stock data, daily prices were averaged across all pharmaceutical companies to create aggregate measures. This approach reduces noise and highlights general trends in the sector. In addition, only the closing daily stock price (averaged across all pharma companies) was used in the analyses. Event data was processed to create dummy variables indicating the occurrence of an event. Each event variable was set to 1 for a 30-day window starting from the event

 emergence beginning date, peak date, and ending date and 0 otherwise.

The comprehensive analysis of the data through these methodologies allows for a thorough understanding of the various factors impacting pharmaceutical stock prices. By segmenting the analysis into different scenarios, the study provides a nuanced view of the influence of market sentiment, macroeconomic indicators, and significant health-related events on the pharmaceutical sector.

\begin{figure*}
\centering
\includegraphics[width=14cm]{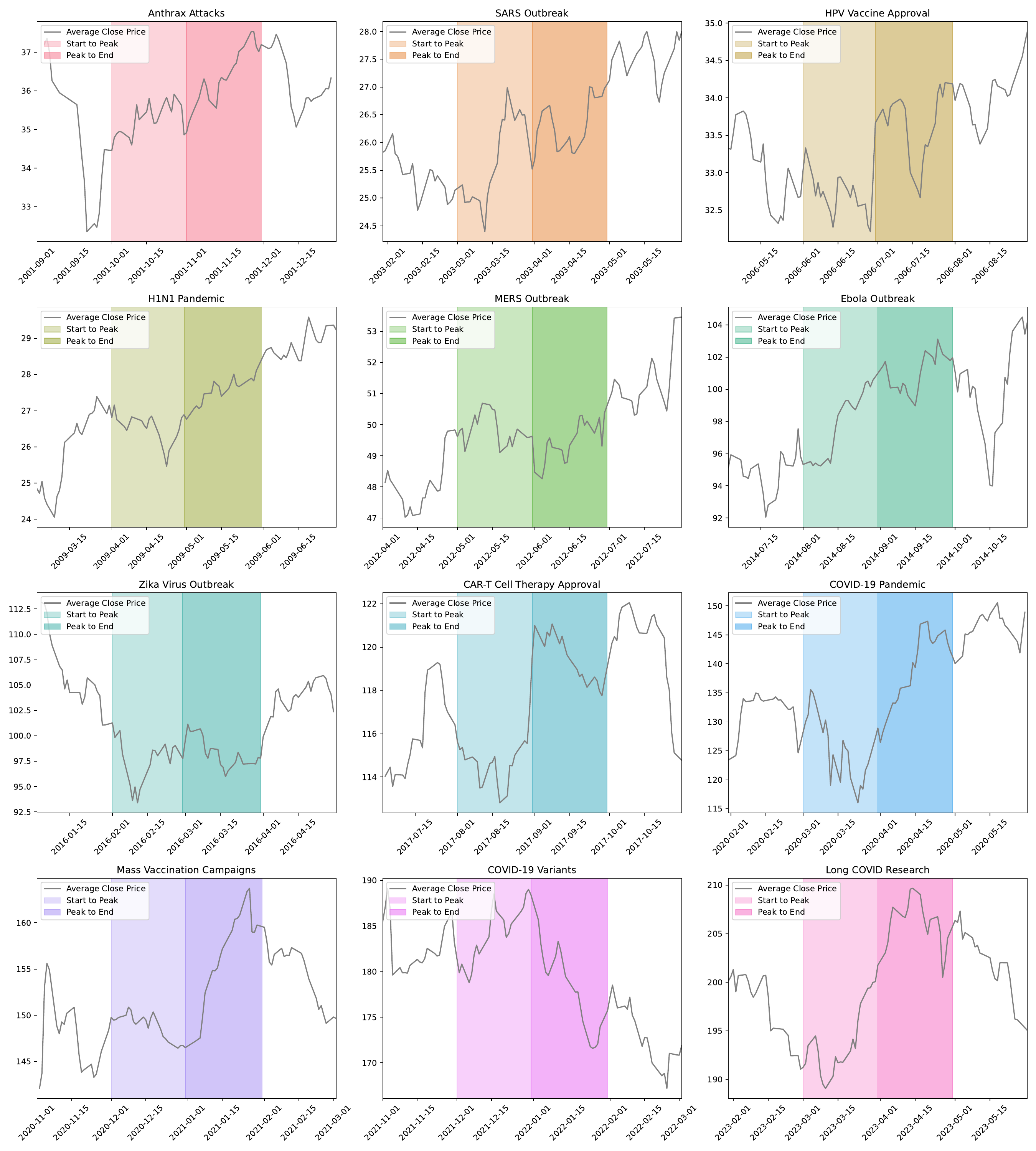}
\caption{This figure consists of 12 subplots, each zooming in on the average closing price of pharmaceutical stocks during the periods of significant health events. Each subplot highlights the start, peak, and end periods of the events with different colors, providing a detailed view of stock price movements during these critical times. The beginning (30 days) and ending phase (30 days) windows are displayed and for simplicity the peak phase (30 days, starting from the middle of beginning phase and running until the middle of ending phase) is removed. \label{fig:peak_subplots}}
\end{figure*}

Figure \ref{fig:peak_subplots} highlights the significant fluctuations in the average closing prices of pharmaceutical stocks in response to major health events over the years. Each subplot corresponds to a specific event, such as the Anthrax attacks, SARS outbreak, HPV vaccine approval, H1N1 pandemic, MERS outbreak, Ebola outbreak, Zika virus outbreak, CAR-T cell therapy approval, COVID-19 pandemic, mass COVID-19 vaccinations, emergence of COVID-19 variants, and long COVID research. By zooming in on the periods surrounding these events, we can observe how stock prices reacted before, during, and after each crisis. For example, the COVID-19 pandemic subplot shows a sharp increase in stock prices as companies involved in vaccine development saw significant gains, followed by fluctuations as the situation evolved. This visualization underscores the direct impact of health crises on the pharmaceutical sector, illustrating how investor confidence and market dynamics shift in response to public health developments. It also highlights the importance of timely market responses and the volatility associated with these events, providing valuable insights for investors and policymakers in managing financial risk during health crises.

The following section will present the results of the regression analyses, highlighting the key findings and their implications for investors, policymakers, and the pharmaceutical industry.

\section{Results}

In this section, we present the findings of our analysis, correlation, and regression results.

\begin{figure*}
\centering
\includegraphics[width=16cm]{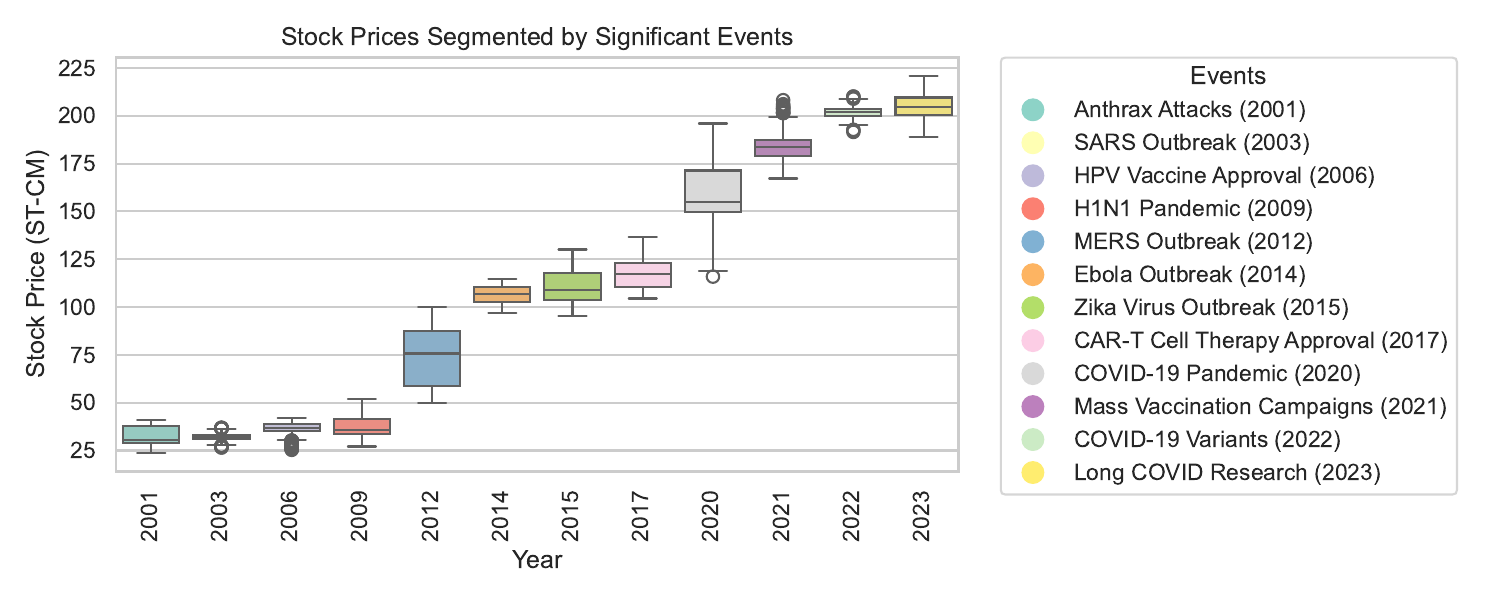}
\caption{Box plots showing the stock prices (ST-CM) segmented by significant health-related events from 2000 to 2023. Each box represents the distribution of stock prices during the specified event period, highlighting the impact of each event on the stock market.\label{fig:boxplot}}
\end{figure*}

Figure \ref{fig:boxplot} displays the increasing trend of daily closing stock price over time, benchmarked by each major health event since 2000. The increase is proven to not be steady with major increases in certain years compared to the previous ones, however, there has been no major decline before, during, or after the events.

\begin{figure*}
\centering
\includegraphics[width=15cm]{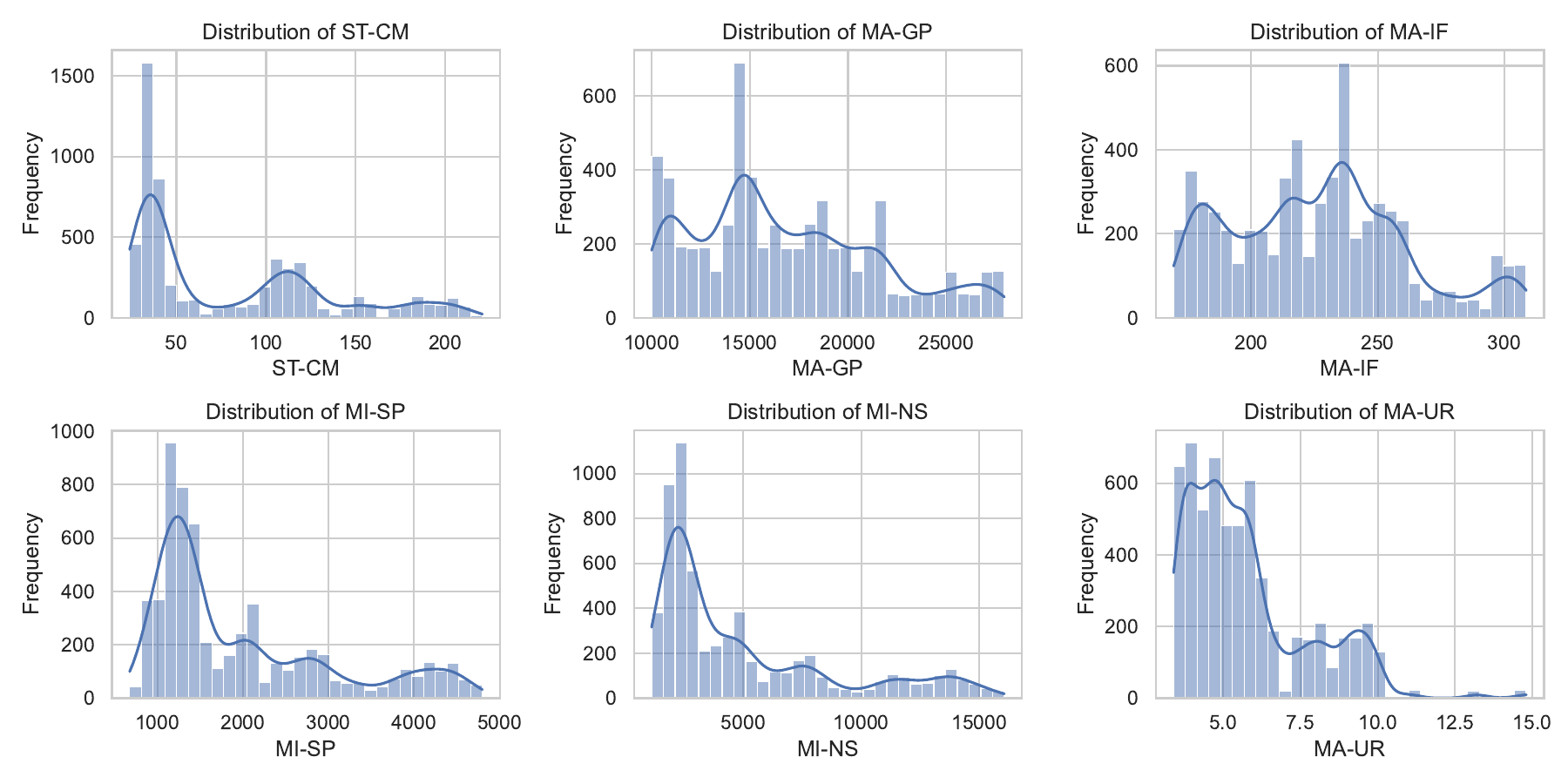}
\caption{This figure presents the histograms of key variables, displaying their distribution over the study period. The variables include ST-CM (Stock Closing Price), MA-GP (Gross Private Domestic Investment), MA-IF (Inflation Rate), MI-SP (S\&P 500 Index), MI-NS (NASDAQ Index), and MA-UR (Unemployment Rate).\label{fig:distributions}}
\end{figure*}

Figure \ref{fig:distributions} provides a visual summary of the distribution of key variables over the study period, offering several insights. The variable ST-CM (Stock Closing Price) displays a right-skewed distribution, suggesting that most closing prices cluster at lower values with fewer instances of higher prices. The MA-GP (Gross Private Domestic Investment) shows multiple peaks, indicating periods of increased and decreased investment activity. The MA-IF (Inflation Rate) histogram reveals a varied distribution, reflecting different inflationary periods over time. The MI-SP (S\&P 500 Index) and MI-NS (NASDAQ Index) histograms illustrate how these major indices have fluctuated, with significant frequencies at certain price levels. Lastly, the MA-UR (Unemployment Rate) distribution shows a range of unemployment levels, capturing economic cycles of growth and recession. Collectively, these histograms help in understanding the underlying patterns and trends in the data, which are crucial for analyzing the impact of macroeconomic and market factors on pharmaceutical stock performance. The insights gained from these distributions can aid in identifying periods of market stability and volatility, thereby providing a comprehensive background for the subsequent regression analysis.

\subsection{Correlation Analysis}

\begin{figure*}
\centering
\includegraphics[width=14cm]{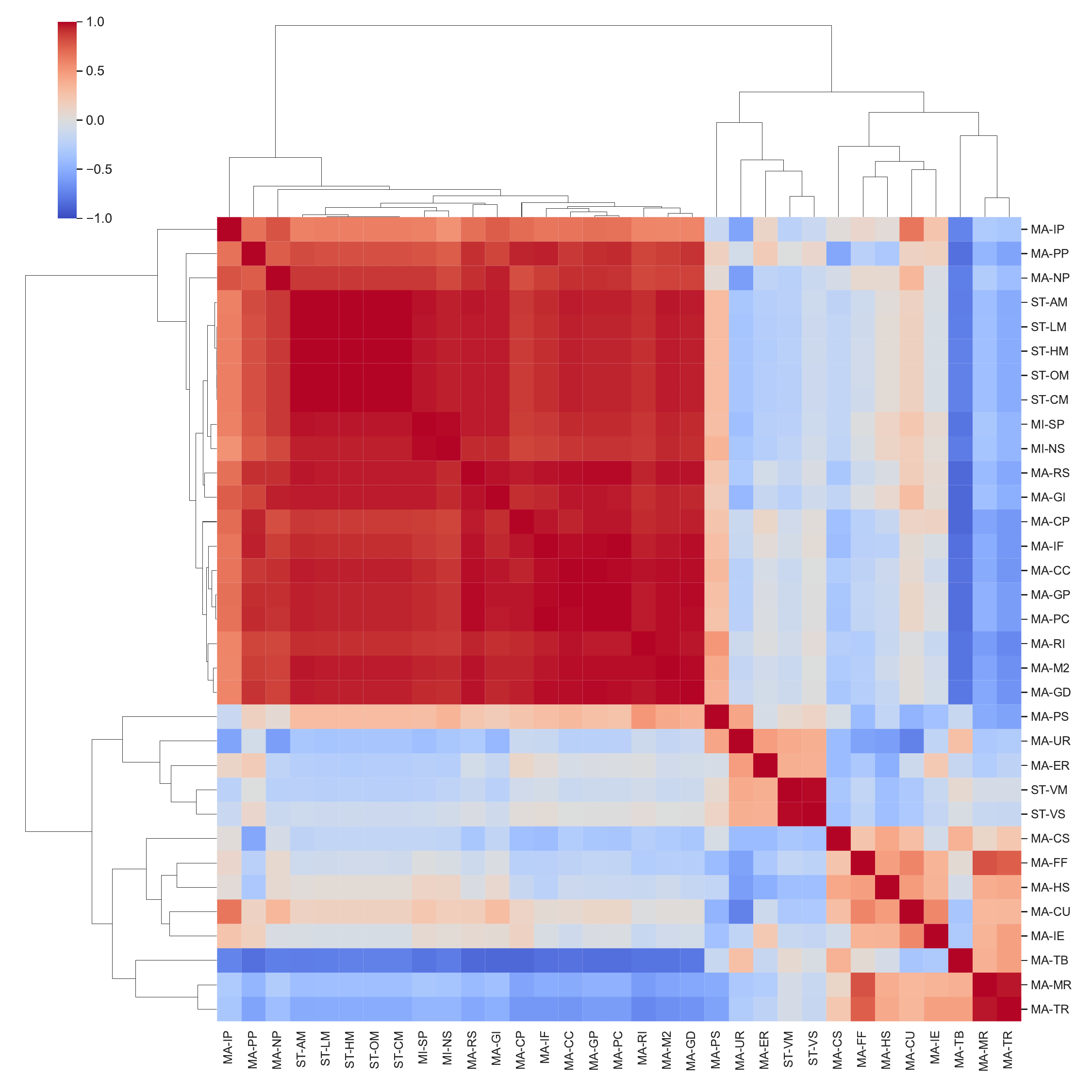}
\caption{Correlation heatmap of various stock market and macroeconomic variables. The heatmap depicts the correlation coefficients between different variables, with red indicating strong positive correlations and blue indicating strong negative correlations. Key variables include stock prices, market indices (S\&P 500, NASDAQ), and macroeconomic indicators such as GDP, inflation rate, and unemployment rate. The hierarchical clustering on both axes groups variables with similar correlation patterns, highlighting the interrelationships among the factors influencing pharmaceutical stock prices.\label{fig:correlation}}
\end{figure*}

Before diving into the regression results, we performed a correlation analysis to understand the relationships between different variables. The correlation heatmap shown in Figure \ref{fig:correlation} provides a visual representation of the strength and direction of these relationships.

The heatmap reveals significant correlations between many variables, particularly within the same category. For instance, stock-related variables (e.g., ST-OM, ST-HM, ST-LM, ST-CM) show high positive correlations with each other, as expected. Similarly, several macroeconomic variables (e.g., MA-GP, MA-IF, MA-UR) are strongly correlated, indicating that they tend to move together. This high correlation suggests potential multicollinearity issues, which were addressed using Ordinary Least Squares (OLS) regression in our analysis.

\subsection{Regression Analysis}

The OLS regression analysis was conducted to examine the impact of various macroeconomic, market, and event variables on the average close price of pharmaceutical stocks (ST-CM). The equations for each model (Models 1 - 4) have been listed in the following:

\begin{equation*}\centering
\text{ST-CM} = \beta_0 + \beta_1 \times \text{(Event Variables)} + \epsilon \label{eq: model 1}
\end{equation*}

\begin{equation*}\centering
\text{ST-CM} = \beta_0 + \beta_1 \times \text{(Event Variables)} + \beta_2 \times \text{(Market Indicators)} + \epsilon \label{eq: model 2}
\end{equation*}

\begin{equation*}\centering
\text{ST-CM} = \beta_0 + \beta_1 \times \text{(Event Variables)} + \beta_2 \times \text{(Macroeconomic Indicators)} + \epsilon \label{eq: model 3}
\end{equation*}

\begin{equation*}
\text{ST-CM} = \beta_0 + \beta_1 \times \text{(Event Variables)} + \beta_2 \times \text{(Market Indicators)} + \beta_3 \times \text{(Macroeconomic Indicators)} + \epsilon \label{eq: model 4}
\end{equation*}

Subsequently, the 4 scenarios outlines in Equations \ref{eq: model 1}, \ref{eq: model 2}, \ref{eq: model 3}, and \ref{eq: model 4} are further illustrated below:

\subsubsection{Scenario 1: No Controls}

In the first scenario, all macroeconomic (MA) and market (MI) variables were included in the model. Several event variables showed significant impacts on pharmaceutical stock prices. For instance, the beginning phase of the CAR-T cell therapy approval (EV-CT.B) had a significantly positive impact on stock prices, reflecting the market's optimism regarding this breakthrough treatment. Conversely, the end phase of the H1N1 pandemic (EV-HP.E) showed a negative impact, likely due to the market's reaction to the prolonged economic and social disruptions caused by the pandemic.

\subsubsection{Scenario 2: Controlling for Market (MI) variables}

In this scenario, we excluded the market indicators

 (MI variables) to focus solely on macroeconomic and event variables. The results show that the exclusion of MI variables does not drastically change the significance of most event variables. The positive impacts of EV-CT.B and the negative impacts of EV-HP.E remain robust, indicating that these events have strong effects on stock prices independent of broader market conditions.

\subsubsection{Scenario 3: Controlling for Macroeconomic (MA) variables}

Excluding the macroeconomic variables (MA variables) while retaining market indicators (MI variables) provides a different perspective. The results indicate that event variables still play a crucial role. The end phase of the COVID-19 pandemic (EV-CP.E) shows a significant positive impact on stock prices, suggesting that the market responded positively to the rollout of vaccines and the subsequent recovery phase.

\subsubsection{Scenario 4: Controlling for Market (MI) and Macroeconomic(MA) variables}

In the final scenario, we excluded both macroeconomic and market variables, leaving only event variables. This approach isolates the direct impact of major health events on pharmaceutical stock prices. The results underscore the significant influence of these events, with several variables showing strong statistical significance. For instance, the end phase of the MERS outbreak (EV-MO.E) had a negative impact on stock prices, highlighting the market's reaction to the prolonged health crisis.

\begin{table*} \centering 
  \caption{This table presents the Ordinary Least Squares (OLS) regression results for different scenarios, analyzing the impact of major health events on pharmaceutical stock prices.} 
  \label{tab:regression} 
\begin{tabular}{@{\extracolsep{5pt}}lcccc} 
\hline  
 & \multicolumn{4}{c}{\textit{Dependent variable: ST-CM}} \\ 
\cline{2-5} 
 & No Controls & Controlling MI & Controlling MA & Controlling MI \& MA \\ 
\hline

\textbf{Beginning} \\
EV-AA.B &  &  & 2.14 (3.43) & \textbf{$-$38.32$^{***}$} (13.89) \\ 
EV-SO.B & 2.25 (1.58) & \textbf{4.73$^{***}$} (1.71) & \textbf{8.03$^{**}$} (3.50) & \textbf{$-$51.00$^{***}$} (14.15) \\ 
EV-HV.B & $-$0.78 (1.52) & \textbf{$-$2.86$^{*}$} (1.65) & \textbf{$-$10.45$^{***}$} (3.45) & \textbf{$-$41.65$^{***}$} (13.96) \\ 
EV-HP.B & 2.45 (1.60) & $-$0.21 (1.73) & \textbf{10.74$^{***}$} (3.57) & \textbf{$-$47.95$^{***}$} (14.41) \\ 
EV-MO.B & $-$1.82 (1.52) & \textbf{$-$2.98$^{*}$} (1.65) & 3.96 (3.42) & \textbf{$-$25.36$^{*}$} (13.89) \\ 
EV-EO.B & $-$1.34 (1.55) & $-$1.36 (1.68) & \textbf{17.52$^{***}$} (3.49) & 21.02 (14.16) \\ 
EV-ZV.B & $-$0.03 (1.53) & \textbf{$-$4.96$^{***}$} (1.65) & \textbf{19.55$^{***}$} (3.42) & 18.87 (13.88) \\ 
EV-CT.B & \textbf{10.26$^{***}$} (1.52) & \textbf{10.78$^{***}$} (1.65) & \textbf{7.92$^{**}$} (3.44) & \textbf{35.90$^{***}$} (13.92) \\ 
EV-CP.B & \textbf{25.53$^{***}$} (1.94) & \textbf{35.98$^{***}$} (2.07) & \textbf{11.99$^{***}$} (3.45) & \textbf{51.14$^{***}$} (13.96) \\ 
EV-MV.B & \textbf{$-$12.38$^{***}$} (1.64) & \textbf{$-$13.67$^{***}$} (1.77) & 0.13 (3.50) & \textbf{70.57$^{***}$} (14.01) \\ 
EV-CV.B & \textbf{$-$6.29$^{***}$} (1.58) & 0.24 (1.69) & \textbf{$-$19.19$^{***}$} (3.47) & \textbf{98.99$^{***}$} (13.93) \\ 
EV-LC.B & $-$2.42 (1.58) & \textbf{$-$6.44$^{***}$} (1.70) & \textbf{8.74$^{**}$} (3.45) & \textbf{109.16$^{***}$} (13.92) \\ 

\textbf{Peak} \\
EV-AA.P &  &  & 0.53 (3.98) & $-$0.37 (16.13) \\ 
EV-SO.P & $-$0.41 (1.75) & 0.45 (1.90) & $-$0.11 (4.02) & 0.61 (16.32) \\ 
EV-HV.P & $-$0.01 (1.75) & $-$0.01 (1.90) & 0.22 (4.03) & 0.03 (16.34) \\ 
EV-HP.P & $-$1.91 (1.76) & $-$0.84 (1.90) & $-$1.15 (4.03) & $-$0.70 (16.36) \\ 
EV-MO.P & $-$0.18 (1.73) & $-$0.87 (1.88) & 1.78 (3.98) & $-$0.64 (16.13) \\ 
EV-EO.P & 0.75 (1.75) & 1.50 (1.90) & $-$0.29 (4.03) & 1.40 (16.34) \\ 
EV-ZV.P & 1.31 (1.83) & 1.59 (1.98) & 1.53 (4.20) & 1.68 (17.03) \\ 
EV-CT.P & 1.80 (1.73) & 1.12 (1.88) & 2.54 (3.99) & 0.77 (16.17) \\ 
EV-CP.P & $-$2.06 (1.78) & \textbf{$-$10.08$^{***}$} (1.91) & \textbf{7.11$^{*}$} (4.03) & $-$10.59 (16.36) \\ 
EV-MV.P & \textbf{$-$5.45$^{***}$} (1.77) & \textbf{$-$5.68$^{***}$} (1.93) & $-$5.09 (4.08) & $-$5.34 (16.54) \\ 
EV-CV.P & \textbf{3.44$^{*}$} (1.76) & \textbf{6.98$^{***}$} (1.90) & $-$1.59 (4.02) & 6.39 (16.32) \\ 
EV-LC.P & \textbf{2.98$^{*}$} (1.74) & 2.41 (1.88) & 4.32 (3.99) & 2.05 (16.17) \\ 

\textbf{Ending} \\
EV-AA.E &  &  & 1.06 (3.42) & \textbf{$-$37.27$^{***}$} (13.87) \\ 
EV-SO.E & 0.98 (1.57) & \textbf{3.67$^{**}$} (1.70)

 & \textbf{6.02$^{*}$} (3.49) & \textbf{$-$50.33$^{***}$} (14.15) \\ 
EV-HV.E & 2.34 (1.52) & 1.26 (1.65) & \textbf{$-$10.37$^{***}$} (3.44) & \textbf{$-$40.44$^{***}$} (13.92) \\ 
EV-HP.E & \textbf{4.79$^{***}$} (1.57) & 0.47 (1.70) & \textbf{8.12$^{**}$} (3.44) & \textbf{$-$46.84$^{***}$} (13.92) \\ 
EV-MO.E & \textbf{$-$3.64$^{**}$} (1.48) & \textbf{$-$4.51$^{***}$} (1.60) & 4.86 (3.32) & \textbf{$-$25.92$^{*}$} (13.45) \\ 
EV-EO.E & $-$1.24 (1.55) & $-$0.09 (1.68) & \textbf{18.98$^{***}$} (3.49) & \textbf{24.48$^{*}$} (14.16) \\ 
EV-ZV.E & $-$0.20 (1.63) & $-$0.68 (1.76) & \textbf{13.08$^{***}$} (3.69) & 19.23 (14.95) \\ 
EV-CT.E & \textbf{11.58$^{***}$} (1.48) & \textbf{11.79$^{***}$} (1.60) & \textbf{11.31$^{***}$} (3.32) & \textbf{40.87$^{***}$} (13.44) \\ 
EV-CP.E & \textbf{9.34$^{***}$} (2.28) & \textbf{33.07$^{***}$} (2.31) & \textbf{22.56$^{***}$} (3.33) & \textbf{63.58$^{***}$} (13.50) \\ 
EV-MV.E & \textbf{3.15$^{*}$} (1.63) & 1.77 (1.77) & 3.29 (3.50) & \textbf{77.02$^{***}$} (13.98) \\ 
EV-CV.E & \textbf{$-$11.48$^{***}$} (1.57) & \textbf{$-$6.78$^{***}$} (1.70) & \textbf{$-$25.68$^{***}$} (3.46) & \textbf{93.48$^{***}$} (13.93) \\ 
EV-LC.E & \textbf{8.88$^{***}$} (1.51) & \textbf{6.72$^{***}$} (1.63) & \textbf{12.53$^{***}$} (3.34) & \textbf{121.46$^{***}$} (13.44) \\ 

Constant & \textbf{76.60$^{***}$} (0.17) & \textbf{74.23$^{***}$} (0.16) & \textbf{79.68$^{***}$} (0.17) & \textbf{78.62$^{***}$} (0.69) \\ 
\hline
R$^{2}$ & 0.99 & 0.99 & 0.95 & 0.10 \\ 
\hline 
\textit{Note:}  & \multicolumn{4}{r}{$^{*}$p$<$0.1; $^{**}$p$<$0.05; $^{***}$p$<$0.01} \\ 
\end{tabular} 
\end{table*}

\section{Discussion}

The regression results provide a comprehensive view of the factors influencing pharmaceutical stock prices under different scenarios. This section delves into the key findings, interpreting the significant variables, and discussing their implications.

\subsection{Impact of major health events}

The analysis reveals that major health events have profound effects on pharmaceutical stock prices. Events such as the SARS outbreak (EV-SO), H1N1 pandemic (EV-HP), and COVID-19 pandemic (EV-CP) show strong and statistically significant impacts. The positive coefficients for these events in the "Beginning" phase indicate initial market optimism and increased demand for pharmaceutical products and services. For instance, the coefficient for COVID-19 in the "Beginning" phase is significantly positive across all models, highlighting the surge in pharmaceutical stocks during the early stages of the pandemic.

Conversely, the "Ending" phase of major health events often shows negative impacts, reflecting market corrections and stabilization after the initial surge. The negative coefficients for COVID-19 in the "Ending" phase indicate that, as the pandemic progressed, the market adjusted to the new normal, leading to a decline in stock prices. This pattern is consistent with previous research that found similar trends during the SARS and H1N1 outbreaks \cite{loh2006impact}.

\subsection{Role of Macroeconomic Variables}

Macroeconomic variables such as GDP (MA-GP), inflation rate (MA-IF), and unemployment rate (MA-UR) also play a crucial role in influencing pharmaceutical stock prices. The results show that higher GDP and lower unemployment rates are positively correlated with stock prices, indicating that strong economic performance boosts investor confidence in the pharmaceutical sector. This finding aligns with previous studies that emphasized the importance of macroeconomic stability for stock market performance \cite{widjaja2023impact}.

Inflation, on the other hand, has a mixed impact. While moderate inflation can indicate economic growth, high inflation rates can erode purchasing power and investor confidence. The negative coefficients for inflation in the models without macroeconomic aggregates (MA) suggest that inflationary pressures adversely affect pharmaceutical stocks when considered in isolation.

\subsection{Market Sentiment and Indices}

Market sentiment, captured through indices such as the S\&P 500 (MI-SP) and NASDAQ (MI-NS), significantly affects pharmaceutical stock prices. The strong positive coefficients for these indices in model 1 and model 4 indicate that broader market performance is a critical determinant of sector-specific stock movements. This finding is consistent with the literature that highlights the sensitivity of pharmaceutical stocks to overall market trends \cite{zhao2022empirical}.

\subsection{Event Data Analysis}

Event data, encompassing significant health-related events, provides critical insights into market reactions. For example, the approval of the HPV vaccine (EV-HV) and CAR-T cell therapy (EV-CT) show strong positive effects on stock prices, reflecting investor optimism about new product launches and potential market expansion. The coefficients for these events are consistently significant across different models, underscoring their importance in driving stock performance.

Interestingly, events like the Ebola outbreak (EV-EO) and Zika virus outbreak (EV-ZV) show varying impacts. While the initial phases of these outbreaks positively influenced stock prices due to increased demand for treatments, the later phases often led to market corrections. This pattern highlights the dynamic nature of market responses to major health events and the importance of timely and effective public health interventions.

\subsection{Comparative Analysis Across Models}

Comparing the results across different scenarios reveals the robustness of the findings. The inclusion or exclusion of specific groups of variables (MI, MA) provides a nuanced understanding of their relative importance. The high R-squared values in models 1, 2, and 3 indicate that these models explain a significant portion of the variance in stock prices. The substantial drop in the R-squared value for model 4, which excludes MI and MA, underscores the critical role these variables play in influencing stock prices.

\subsection{Policy Implications and Investment Strategies}

The findings have important policy implications and can inform investment strategies. Policymakers should recognize the significant impact of macroeconomic stability and effective crisis management on the pharmaceutical sector. Ensuring a stable economic environment and swift response to major health events can bolster investor confidence and stabilize stock markets.

For investors, understanding the interplay between macroeconomic variables, market sentiment, and major health events is crucial for making informed investment decisions. The results suggest that periods of economic growth and positive market sentiment present lucrative opportunities for investing in pharmaceutical stocks. Additionally, being attuned to significant health-related events and their potential market impacts can help investors capitalize on short-term gains while mitigating risks associated with market corrections.

\section{Conclusion}

This study provides a comprehensive analysis of the factors influencing pharmaceutical stock prices, highlighting the significant impacts of macroeconomic variables, market sentiment, and major health events. The findings underscore the dynamic nature of the pharmaceutical sector and the importance of considering a wide range of variables when analyzing stock performance. The regression results reveal that major health events, such as

 the SARS, H1N1, and COVID-19 pandemics, have profound and multifaceted impacts on pharmaceutical stocks. The initial phases of these crises often lead to market optimism and increased demand for pharmaceutical products, while the later phases see market corrections as conditions stabilize. Macroeconomic variables, including GDP, inflation, and unemployment rates, play a crucial role in shaping investor perceptions and stock performance. Strong economic performance and low unemployment rates boost investor confidence, whereas inflationary pressures can have mixed effects.

Market sentiment, captured through indices like the S\&P 500 and NASDAQ, is a critical determinant of pharmaceutical stock movements. Broader market trends significantly influence sector-specific stock prices, underscoring the importance of considering overall market conditions in investment strategies. Event data analysis highlights the market's dynamic responses to significant health-related events. The approval of new treatments and vaccines, as well as outbreaks of infectious diseases, drive substantial fluctuations in stock prices. These findings emphasize the importance of timely public health interventions and effective communication strategies.

In conclusion, this study offers valuable insights into the factors affecting pharmaceutical stock prices, providing a robust foundation for future research and practical applications. Policymakers and investors alike can benefit from understanding the complex interplay of macroeconomic conditions, market sentiment, and major health events in shaping stock market dynamics.

\section{Data Availability}

The data and code used in this analysis is provided in the following GitHub repository: \href{https://github.com/mprtrmrtz/health_crises_pharma_stocks}{Major Health Events Impacts on Pharma Stocks} 

\section{Acknowledgments}
We highly appreciate the data providers at Yahoo Finance and Federal Reserve Bank for providing publicly available financial data.

\section{Abbreviations}
The following abbreviations are used in this manuscript:\\

\noindent 
\begin{tabular}{@{}ll}
SARS & Severe Acute Respiratory Syndrome\\
H1N1 & Influenza A virus subtype H1N1\\
MERS & Middle East Respiratory Syndrome\\
COVID-19 & Coronavirus Disease 2019\\
AI & Artificial Intelligence\\
GDP & Gross Domestic Product\\
FDA & Food and Drug Administration\\
WHO & World Health Organization\\
CDC & Centers for Disease Control and Prevention\\
OLS & Ordinary Least Squares\\
SP & S\&P 500 Index\\
NASDAQ & National Association of Securities Dealers Automated Quotations\\
ST & Stock Data\\
MI & Market Data \\
MA & Macroeconomic Data\\
\end{tabular}

\bibliographystyle{plainnat}
\bibliography{references}

\begin{thebibliography}{34}
\providecommand{\natexlab}[1]{#1}
\providecommand{\url}[1]{\texttt{#1}}
\expandafter\ifx\csname urlstyle\endcsname\relax
  \providecommand{\doi}[1]{doi: #1}\else
  \providecommand{\doi}{doi: \begingroup \urlstyle{rm}\Url}\fi

\bibitem[Akaev and Sadovnichiy(2021)]{akaev2021information}
Askar Akaev and Viktor Sadovnichiy.
\newblock Information models for forecasting nonlinear economic dynamics in the
  digital era.
\newblock \emph{Applied Mathematics}, 12\penalty0 (3):\penalty0 171--208, 2021.

\bibitem[Al-Awadhi et~al.(2020)Al-Awadhi, Alsaifi, Al-Awadhi, and
  Alhammadi]{al-awadhi2020}
Abdullah~M. Al-Awadhi, Khaled Alsaifi, Abdullah Al-Awadhi, and Salah Alhammadi.
\newblock Death and contagious infectious diseases: Impact of the covid-19
  virus on stock market returns.
\newblock \emph{Journal of Behavioral and Experimental Finance}, 27:\penalty0
  100326, 2020.
\newblock \doi{10.1016/j.jbef.2020.100326}.
\newblock URL \url{https://doi.org/10.1016/j.jbef.2020.100326}.

\bibitem[Ashraf(2020)]{ashraf2020}
Badar~Nadeem Ashraf.
\newblock Stock markets’ reaction to covid-19: Cases or fatalities?
\newblock \emph{Research in International Business and Finance}, 54:\penalty0
  101249, 2020.
\newblock \doi{10.1016/j.ribaf.2020.101249}.
\newblock URL \url{https://doi.org/10.1016/j.ribaf.2020.101249}.

\bibitem[Barro and Urs{\'u}a(2008)]{barro2008macroeconomic}
Robert~J Barro and Jos{\'e}~F Urs{\'u}a.
\newblock Macroeconomic crises since 1870.
\newblock Technical report, National Bureau of Economic Research, 2008.

\bibitem[Borio et~al.(2019)Borio, Drehmann, and Xia]{borio2019predicting}
Claudio~EV Borio, Mathias Drehmann, and Fan~Dora Xia.
\newblock Predicting recessions: financial cycle versus term spread.
\newblock \emph{BIS working paper}, 2019.

\bibitem[{Centers for Disease Control and Prevention (CDC)}(2024)]{cdc}
{Centers for Disease Control and Prevention (CDC)}.
\newblock Cdc data and statistics.
\newblock \url{https://www.cdc.gov/}, 2024.
\newblock Accessed: 2024-07-02.

\bibitem[Chen et~al.(2009)Chen, Chen, Tang, and Huang]{chen2009positive}
Chun-Da Chen, Chin-Chun Chen, Wan-Wei Tang, and Bor-Yi Huang.
\newblock The positive and negative impacts of the sars outbreak: A case of the
  taiwan industries.
\newblock \emph{The Journal of Developing Areas}, pages 281--293, 2009.

\bibitem[Danielsson et~al.(2018)Danielsson, Valenzuela, and
  Zer]{danielsson2018learning}
Jon Danielsson, Marcela Valenzuela, and Ilknur Zer.
\newblock Learning from history: Volatility and financial crises.
\newblock \emph{The Review of Financial Studies}, 31\penalty0 (7):\penalty0
  2774--2805, 2018.

\bibitem[{Federal Reserve Bank of Saint Louis}(2024)]{FRED_API}
{Federal Reserve Bank of Saint Louis}.
\newblock Fred api, 2024.
\newblock URL \url{https://fred.stlouisfed.org/}.
\newblock Accessed: 2024-07-02.

\bibitem[Hoerl and Kennard(1970)]{hoerl1970ridge}
Arthur~E Hoerl and Robert~W Kennard.
\newblock Ridge regression: Biased estimation for nonorthogonal problems.
\newblock \emph{Technometrics}, 12\penalty0 (1):\penalty0 55--67, 1970.

\bibitem[Hunjra et~al.(2014)Hunjra, Chani, Ijaz, and Farooq]{hunjra2014impact}
Ahmed~Imran Hunjra, Dr~Muhammad~Irfan Chani, Muhammad~Shahzad Ijaz, and
  Muhammad Farooq.
\newblock The impact of macroeconomic variables on stock prices in pakistan.
\newblock \emph{International Journal of Economics and Empirical Research},
  2\penalty0 (1):\penalty0 13--21, 2014.

\bibitem[Ibrahim(1999)]{ibrahim1999macroeconomic}
Mansor Ibrahim.
\newblock Macroeconomic variables and stock prices in malaysia: An empirical
  analysis.
\newblock \emph{Asian Economic Journal}, 13\penalty0 (2):\penalty0 219--231,
  1999.

\bibitem[Karamehic et~al.(2013)Karamehic, Ridic, Ridic, Jukic, Coric, Subasic,
  Panjeta, Saban, Zunic, and Masic]{karamehic2013financial}
Jasenko Karamehic, Ognjen Ridic, Goran Ridic, Tomislav Jukic, Jozo Coric, Djemo
  Subasic, Mirsad Panjeta, Aida Saban, Lejla Zunic, and Izet Masic.
\newblock Financial aspects and the future of the pharmaceutical industry in
  the united states of america.
\newblock \emph{Materia socio-medica}, 25\penalty0 (4):\penalty0 286, 2013.

\bibitem[Kumar and Thenmozhi(2006)]{kumar2006forecasting}
Manish Kumar and M~Thenmozhi.
\newblock Forecasting stock index movement: A comparison of support vector
  machines and random forest.
\newblock In \emph{Indian institute of capital markets 9th capital markets
  conference paper}, 2006.

\bibitem[Loh(2006)]{loh2006impact}
Elaine Loh.
\newblock The impact of sars on the performance and risk profile of airline
  stocks.
\newblock \emph{International journal of transport economics: Rivista
  internazionale di economia dei trasporti: XXXIII, 3, 2006}, pages 1000--1022,
  2006.

\bibitem[Maleki and Ghahari(2024)]{healthcare12151458}
Morteza Maleki and SeyedAli Ghahari.
\newblock Comprehensive clustering analysis and profiling of covid-19 vaccine
  hesitancy and related factors across u.s. counties: Insights for future
  pandemic responses.
\newblock \emph{Healthcare}, 12\penalty0 (15), 2024.
\newblock ISSN 2227-9032.
\newblock \doi{10.3390/healthcare12151458}.
\newblock URL \url{https://www.mdpi.com/2227-9032/12/15/1458}.

\bibitem[Maleki et~al.(2022)Maleki, Bahrami, Menendez, and
  Balsa-Barreiro]{maleki2022social}
Morteza Maleki, Mohsen Bahrami, Monica Menendez, and Jose Balsa-Barreiro.
\newblock Social behavior and covid-19: Analysis of the social factors behind
  compliance with interventions across the united states.
\newblock \emph{International journal of environmental research and public
  health}, 19\penalty0 (23):\penalty0 15716, 2022.

\bibitem[Milosevic(2016)]{milosevic2016equity}
Nikola Milosevic.
\newblock Equity forecast: Predicting long term stock price movement using
  machine learning.
\newblock \emph{arXiv preprint arXiv:1603.00751}, 2016.

\bibitem[Mironiuc et~al.(2022)Mironiuc, Huian, Țaran, and
  Curea]{mironiuc2022financial}
Marilena Mironiuc, Maria~Carmen Huian, Alina Țaran, and Mihaela Curea.
\newblock Financial market reaction to r\&d volatility in the pharmaceutical
  industry. a multi-country study.
\newblock \emph{Journal of Business Economics and Management}, 23\penalty0
  (5):\penalty0 1234--1256, 2022.

\bibitem[Otoo(1999)]{otoo1999consumer}
Maria~Ward Otoo.
\newblock Consumer sentiment and the stock market.
\newblock \emph{Available at SSRN 205028}, 1999.

\bibitem[Siu and Wong(2004)]{siu2004economic}
Alan Siu and YC~Richard Wong.
\newblock Economic impact of sars: The case of hong kong.
\newblock \emph{Asian Economic Papers}, 3\penalty0 (1):\penalty0 62--83, 2004.

\bibitem[Tibshirani(1996)]{tibshirani1996lasso}
Robert Tibshirani.
\newblock Regression shrinkage and selection via the lasso.
\newblock \emph{Journal of the Royal Statistical Society Series B: Statistical
  Methodology}, 58\penalty0 (1):\penalty0 267--288, 1996.

\bibitem[Tsai(2015)]{tsai2015us}
Chun-Li Tsai.
\newblock How do us stock returns respond differently to oil price shocks
  pre-crisis, within the financial crisis, and post-crisis?
\newblock \emph{Energy Economics}, 50:\penalty0 47--62, 2015.

\bibitem[{U.S. Food and Drug Administration (FDA)}(2024)]{fda}
{U.S. Food and Drug Administration (FDA)}.
\newblock Fda data and statistics.
\newblock \url{https://www.fda.gov/}, 2024.
\newblock Accessed: 2024-07-02.

\bibitem[Vel{\'a}squez et~al.(2022)Vel{\'a}squez, Gri{\~n}en, and
  Henr{\'\i}quez]{velasquez2022emerging}
Jorge~Sep{\'u}lveda Vel{\'a}squez, Pablo~Tapia Gri{\~n}en, and Boris~Past{\'e}n
  Henr{\'\i}quez.
\newblock Emerging market dynamics in h1n1 and covid-19 pandemics.
\newblock \emph{Economics Letters}, 218:\penalty0 110766, 2022.

\bibitem[Wang and Liu(2022)]{wang2022pandemic}
Qiuyun Wang and Lu~Liu.
\newblock Pandemic or panic? a firm-level study on the psychological and
  industrial impacts of covid-19 on the chinese stock market.
\newblock \emph{Financial Innovation}, 8\penalty0 (1):\penalty0 36, 2022.

\bibitem[Widjaja et~al.(2023)Widjaja, Sembel, and Malau]{widjaja2023impact}
Weki Widjaja, Roy Sembel, and Melinda Malau.
\newblock The impact of macroeconomic, financial performance, market return,
  and covid-19 instances on stock return of pharmaceutical companies.
\newblock \emph{South East Asia Journal of Contemporary Business, Economics and
  Law}, 28\penalty0 (3):\penalty0 48--57, 2023.

\bibitem[{World Health Organization (WHO)}(2024)]{who}
{World Health Organization (WHO)}.
\newblock Who data and statistics.
\newblock \url{https://www.who.int/}, 2024.
\newblock Accessed: 2024-07-02.

\bibitem[Wu and Chong(2021)]{wu2021does}
Zhang Wu and Terence Tai-Leung Chong.
\newblock Does the macroeconomy matter to market volatility? evidence from us
  industries.
\newblock \emph{Empirical Economics}, 61\penalty0 (6):\penalty0 2931--2962,
  2021.

\bibitem[{Yahoo Finance}(2024{\natexlab{a}})]{yahoo_finance}
{Yahoo Finance}.
\newblock Financial data and statistics.
\newblock \url{https://finance.yahoo.com/}, 2024{\natexlab{a}}.
\newblock Accessed: 2024-07-02.

\bibitem[{Yahoo Finance}(2024{\natexlab{b}})]{yahoo_nasdaq}
{Yahoo Finance}.
\newblock Nasdaq composite index.
\newblock \url{https://finance.yahoo.com/quote/NQ=F/}, 2024{\natexlab{b}}.
\newblock Accessed: 2024-07-02.

\bibitem[{Yahoo Finance}(2024{\natexlab{c}})]{yahoo_sp500}
{Yahoo Finance}.
\newblock S\&p 500 index.
\newblock \url{https://finance.yahoo.com/quote/%5EGSPC/}, 2024{\natexlab{c}}.
\newblock Accessed: 2024-07-02.

\bibitem[Zhang et~al.(2020)Zhang, Hu, and Ji]{zhang2020}
Dayong Zhang, Min Hu, and Qiang Ji.
\newblock Financial markets under the global pandemic of covid-19.
\newblock \emph{Finance Research Letters}, 36:\penalty0 101528, 2020.
\newblock \doi{10.1016/j.frl.2020.101528}.
\newblock URL \url{https://doi.org/10.1016/j.frl.2020.101528}.

\bibitem[Zhao(2022)]{zhao2022empirical}
Zihao Zhao.
\newblock Empirical research and regulatory reflections on the us stocks and
  pharmaceutical industry under the covid-19 pandemic.
\newblock In \emph{2022 4th International Conference on Economic Management and
  Cultural Industry (ICEMCI 2022)}, pages 1808--1817. Atlantis Press, 2022.

\end{thebibliography}
\end{document}